\documentclass[useAMS,usenatbib]{mn2e}
\usepackage{amssymb,graphicx,natbib,longtable,hyperref}
\usepackage{breakurl}
\bibpunct{(}{)}{;}{a}{}{,} 

\def \sw {{\em Swift}}

\def \rxs {1RXS J194211.9+255552}
\title[Study on \rxs]{Spectral and timing characterization of the X-ray source \rxs}
\author[]{A. D'A\`i$^{1}$,\thanks{E-mail:antonino.dai@ifc.inaf.it} 
G.Cusumano$^{1}$,
V. La Parola$^{1}$, 
A. Segreto$^{1}$ \\
$^{1}$ INAF - Istituto di Astrofisica Spaziale e Fisica Cosmica, Via U.\ La Malfa 153, I-90146 Palermo, Italy
}

\begin{document}

%\date{Accepted ... Received ...; in original form ...}

\pagerange{\pageref{firstpage}--\pageref{lastpage}} \pubyear{}

\maketitle

\label{firstpage}

\begin{abstract}
We report  on the  first spectral and  timing characterization  of the
transient X-ray source \rxs~ using all available data from the $Swift$
X-ray satellite.  We  used 10 years of hard X-ray  data from the Burst
Alert Telescope, to characterize its long-term behaviour and to search
for long periodicities, finding evidence  for a periodic modulation at
166.5\,$\pm$\,0.5 days, that we interpret as the orbital period of the
source.  The folded  light curve  reveals that  the X-ray  emission is
mostly concentrated in  a restricted phase-interval and  we propose to
associate \rxs~  to the class of  the Be X-ray binaries.  This is also
supported by the  results of the spectral analysis, where  we used the
BAT data  and three  pointed $Swift$/XRT observations  to characterize
the X-ray broadband spectral shape. We found mild spectral variability
in soft  X-rays that can be  accounted for by a  varying local neutral
absorber,  while the  intrinsic  emission is  consistent  with a  hard
power-law multiplied by a high-energy exponential cut-off as typically
observed in this class of systems.
\end{abstract}

\begin{keywords}
X-rays: binaries -- X-rays: individual: \rxs.
\noindent
Facility: {\it Swift}
\end{keywords}

\section{Introduction} \label{introduction}
The X-ray source  \rxs~ (J194211 hereafter) was first  reported in the
ROSAT                All-Sky               Faint                Source
Catalogue\footnote{\url{http://www.xray.mpe.mpg.de/rosat/survey/rass-fsc/}},
at   RA\,=\,295.54959   and   Dec.\,=\,25.93125  with   a   positional
uncertainty    of   18    arcsec   and    at   a    flux   level    of
(1.8\,$\pm$\,0.7)\,$\times$\,10$^{-13}$   erg    cm$^{-2}$   s$^{-1}$.
Probably due to its relative faintness, the source has been thereafter
poorly  studied.  At  the end  of December  2011, J194211  was clearly
detected during the $INTEGRAL$ Galactic  Plane Scanning in soft X-rays
at 26\,$\pm$\,7 mCrab  \citep{chenevez11atel}. A follow-up observation
with  $Swift$/XRT allowed  for the  first determination  of the  X-ray
spectrum, consistent with an  absorbed (Galactic equivalent absorption
column  1.3\,$\pm$\,0.6\,$\times$\,10$^{22}$  cm$^{-2}$) power-law  of
photon-index $\Gamma$\,=\,0.64 and an unabsorbed flux in the 1--10 keV
range   of    7.88\,$\times$\,10$^{-11}$   erg    cm$^{-2}$   s$^{-1}$
\citep{sidoli11atel}.   The  source  positional   error  lead  to  the
identification  of  two  possible  near  infra-red  counterparts,  and
\citet{sidoli11atel} proposed  the association with the  brighter one,
the 2MASS star 19421116+2556056,  whose optical spectrum (1159-0412038
in  the USNO-B1.0  catalogue) is  consistent with  an early-type  star
\citep{masetti12atel} pointing to the  possible high-mass X-ray binary
nature of the X-ray source.

We present  here the first complete  spectral and timing study  in the
X-ray  domain of  this source  using all  available data  collected by
$Swift$.

	%%%%%%%%%%%%%%%%%%%%%%%%%%%%%%%%%%%%%%%%%%%%%%%%%%%%%%%%%
	\section{Data reduction\label{data}}
	%%%%%%%%%%%%%%%%%%%%%%%%%%%%%%%%%%%%%%%%%%%%%%%%%%%%%%%%%

The survey  data collected  with $Swift$/BAT  between December  2004 and
June    2014    were    retrieved     from    the    HEASARC    public
archive\footnote{\url{http://heasarc.gsfc.nasa.gov/docs/archive.html}}
and processed using  a software dedicated to the analysis  of the data
from coded mask telescopes  \citep{segreto10}.  The source is detected
with  a significance  of 5.8  standard deviations  in the  15--150 keV
energy band, but the signal is optimized in the 15--45 keV energy band
where we obtained a significance of 7.0 standard deviations.

\begin{figure}
\begin{center}
 \includegraphics[width=\columnwidth]{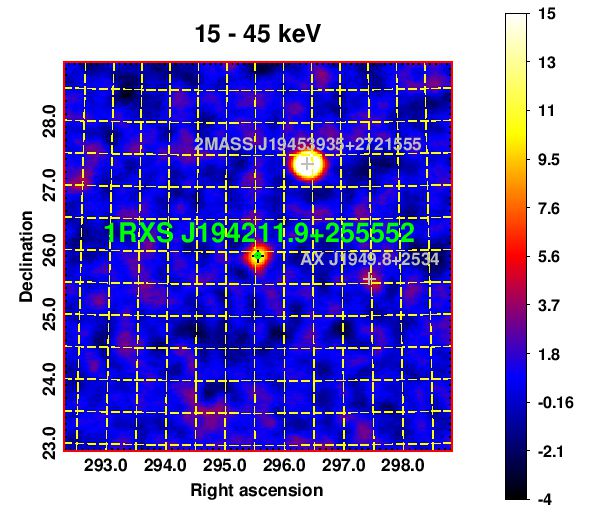} 
\caption[\rxs sky map]{15--45 keV $Swift$/BAT significance map centred on \rxs.}
\label{map}
\end{center}
\end{figure}

\begin{figure}
\begin{center}
\centerline{\includegraphics[height=\columnwidth, angle=-90]{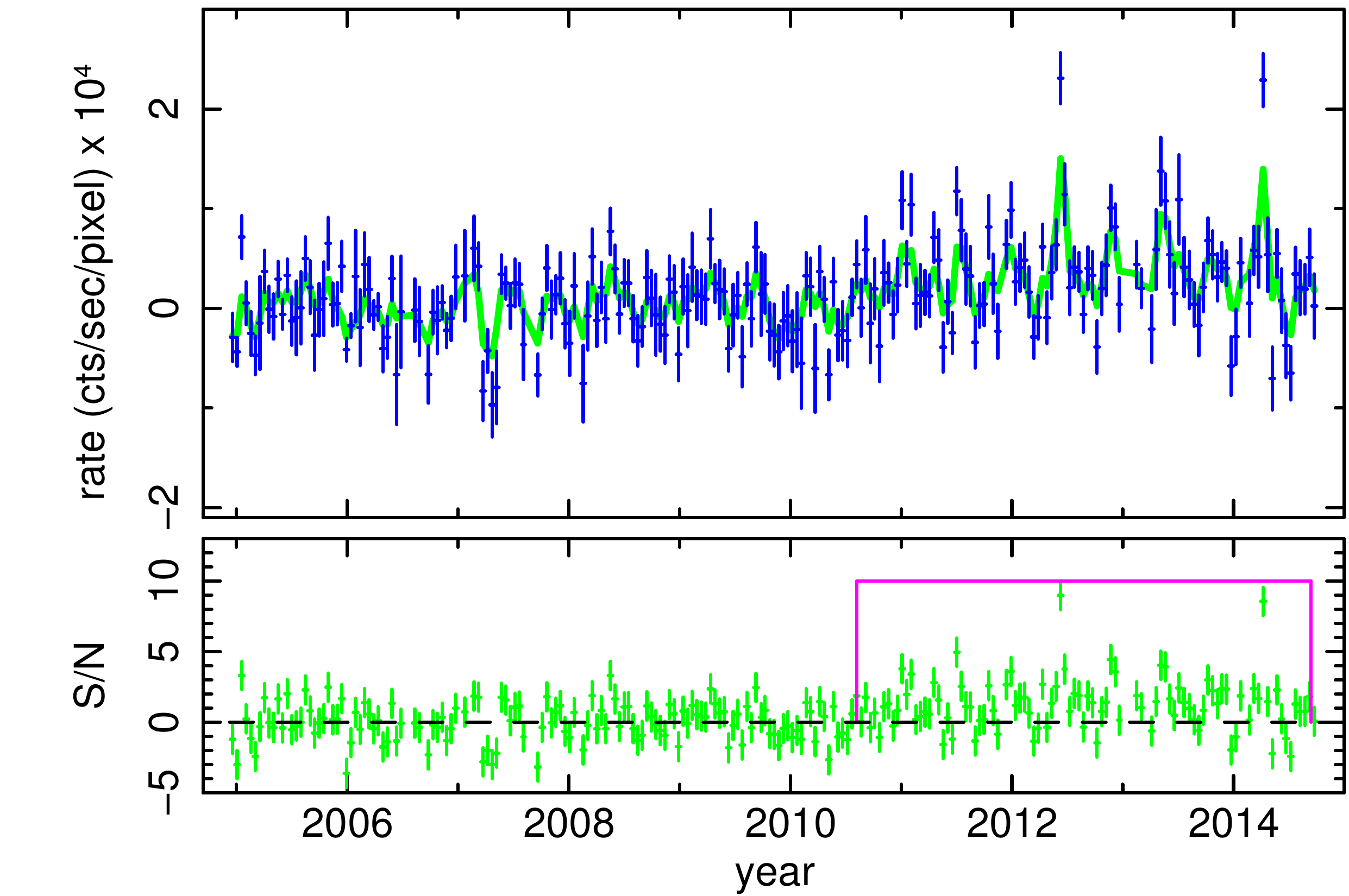}} 
\caption{Long-term 15--45  keV $Swift$/BAT  light curve  (upper panel)
  and  signal-to-noise  ratio  (lower  panel). Bin-time  is  15  days.
  Magenta box indicates the time window of the active X-ray state.}
\label{lc_bat}
\end{center}
\end{figure}

Fig.~\ref{map} shows the significance map of the BAT sky region around
J194211 in  this energy  band.  We  extracted a  background subtracted
light curve  in the 15--45 keV  band, binned at 15-days  resolution to
check   the    long-term   source    variability   in    hard   X-rays
(Fig.~\ref{lc_bat}).  The light curve clearly shows two different flux
levels, with  a significant higher flux  in the more recent  years. To
have an  estimate of the  time when  this enhanced activity  began, we
built the source signal-to-noise ratio (SNR) as a function of time; we
started calculating  the source significance using  the entire 10-year
time-span; we  then progressively  removed one week  of data  from the
start  of the  BAT  monitoring and  obtained the  plot  of the  source
significance as a function of the time window. We obtained the highest
significance using the time window  from 2010.6 to 2014.5 (magenta box
in Fig.~\ref{lc_bat}).  The average BAT rate before 2010.6 has only an
upper limit  of 3.6\,$\times$ 10$^{-6}$ counts  s$^{-1}$ pixel$^{-1}$,
that is compatible with zero, whereas in the period 2010.6--2014.5 the
average rate increases to (3.3\,$\pm$\,0.4)\,$\times$ 10$^{-5}$ counts
s$^{-1}$ pixel$^{-1}$.

During this active X-ray state, the light curve suggested a pattern of
recurrent    peaks    (smoothed    interpolated    green    line    in
Fig.~\ref{lc_bat}),  that   we  investigated  performing   a  temporal
analysis on the data (see Sect.\,3).

\begin{figure}
\begin{center}
\centerline{\includegraphics[width=\columnwidth,height=0.5\columnwidth]{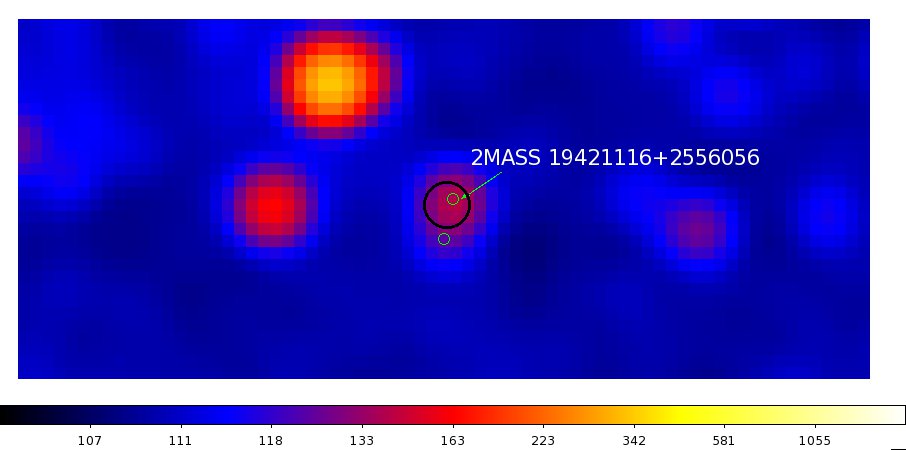}} 
\caption{2MASS sky  region around J194221.  Black  circle: our refined
  $Swift$/XRT  position. Green  circles:  the  two proposed  infra-red
  counterparts as reported in \citet{sidoli11atel}.}
\label{2mass}
\end{center}
\end{figure} 

J194221 was  observed during  three visits  (Obs.ID 00032228001/02/03)
with $Swift$/XRT in Photon Counting mode  at the end of December 2011,
for a total collecting time of 6455 s. In Table~\ref{log}, we show the
log containing the details about the three $Swift$/XRT pointings.

\begin{table*}
\begin{tabular}{r llll}
\hline
Obs.ID            	& Tstart & Tstop & Exposure & Rate    \\ 
                     & UTC    & UTC  & s        & counts s$^{-1}$\\
\hline \hline
00032228001             	& 2011-12-21 06:09:04 & 2011-12-21 07:37:02 & 1955 & 0.5261 \\
00032228002			& 2011-12-24 04:40:04 & 2011-12-24 08:03:56 & 2063 & 0.4602 \\
00032228003			& 2011-12-26 00:03:19 & 2011-12-26 03:25:57 & 2437 & 0.2641\\
\hline
\end{tabular}
\caption{Log of the pointed $Swift$/XRT observations of \rxs.
\label{log} }
\end{table*}

The $Swift$/XRT  data were  processed using  the {\sc  ftools} package
with standard  procedures ({\sc xrtpipeline} v.0.12.4),  filtering and
screening criteria, with standard grade  filtering 0-12. The source is
clearly detected in the three visits and no other contaminating source
is  present in  the  $Swift$/XRT  field of  view  (see  left panel  of
Fig.~\ref{map}). To  better constrain the source  position, we stacked
the  three  images  and  we  obtained a  refined  source  position  at
RA\,=\,19:42:11.22 and Dec.\,=\,+25:56:05.2 (1.9  arcsec error at 90\%
confidence   level)  using   the  online   $Swift$/XRT  build   products
tool\footnote{\url{http://www.swift.ac.uk/user_objects/}}, that allows
also     astrometric    and     enhanced    position     determination
\citep{goad07,evans14}.    This  new   position   and  error   results
consistent with early estimates  \citep{sidoli11atel}, but the smaller
uncertainty  allows  us  to  better support  the  association  of  the
possible infra-red counterpart with the 2MASS object 19421116+2556056,
as shown in Fig.\ref{2mass}.

A first inspection  of the light curves showed  mild variability, with
averaged count  rates per snapshot varying  up to a factor  of 2.  The
maximum count rate is $\sim$\,0.53 c/s in Obs.ID 00032228001, close to
the   threshold    indicated   for   a    pile-up   check\footnote{see
  e.g.    \url{http://www.swift.ac.uk/analysis/xrt/pileup.php}}.    We
verified          according          to         the          suggested
pipeline\footnote{\url{http://www.swift.ac.uk/analysis/xrt/pileup.php}}
that for this  Obs.ID a circular region with 4  pixels radius (centred
at the  source position)  must be excluded  to avoid  spectral pile-up
distortions.   Therefore, the  source spectrum  was extracted  from an
annular  region of  4  and  40 pixels  internal  and external  radius,
respectively,  centred   on  the  source  coordinates   as  previously
determined.   A  background spectrum  was  extracted  from an  annular
region of inner and outer radius  90 and 150 pixels, respectively.  No
pile-up correction was  found necessary for the other  two Obs.ID, and
we used a  circular regions of 40 pixels radius  to extract the source
spectrum and the same annular region for the background spectrum.

The XRT ancillary response file generated with {\sc xrtmkarf} accounts
for PSF and vignetting correction; we used the spectral redistribution
matrix  v013  available  in  the $Swift$  calibration  database.   The
spectral  analysis  was  performed  using {\sc  xspec}  v.12.5,  after
grouping the spectrum with a minimum of 20 counts per channel to allow
the use of $\chi^2$ statistics.   The source events arrival times were
corrected     to      the     SSB     using     the      task     {\sc
  barycorr}\footnote{\url{http://http://heasarc.gsfc.nasa.gov/ftools/caldb/help/barycorr.html}}.

        %%%%%%%%%%%%%%%%%%%%%%%%%%%%%%%%%%%%%%%%%%%%%%%%%%%%%%%%%
        \section{Timing analysis} \label{timing}
        %%%%%%%%%%%%%%%%%%%%%%%%%%%%%%%%%%%%%%%%%%%%%%%%%%%%%%%%%

\begin{figure}%%%%%%%%%%%%%%%%%%%%%%%%%%%%%%%%%%%%%%%%%%%%%%%%%%%%% PAP VII FIGURE 2
\begin{center}
\centerline{\includegraphics[width=5.2cm,angle=270]{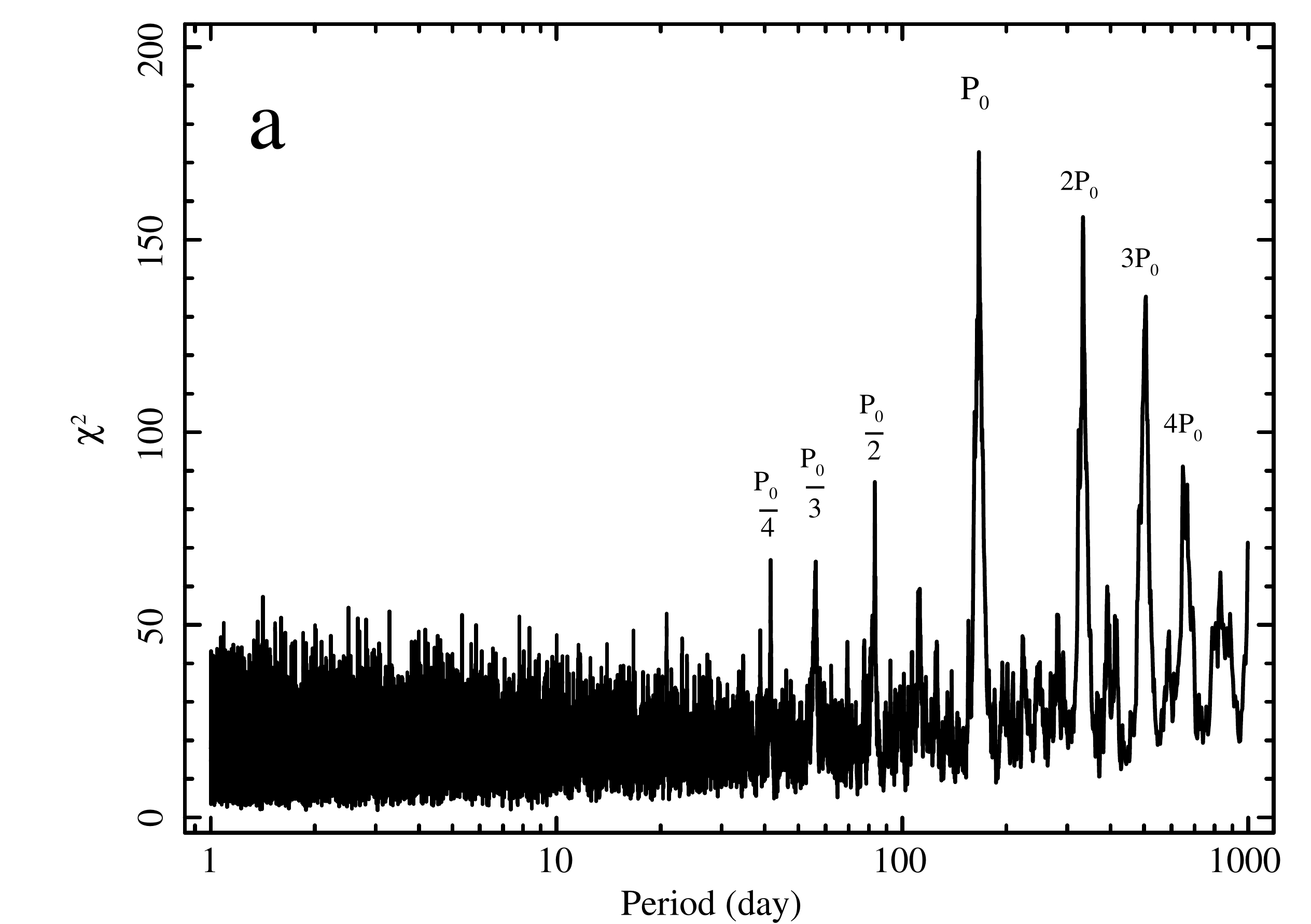}}
%\vspace{0.5truecm}
\centerline{\includegraphics[width=5.2cm,angle=270]{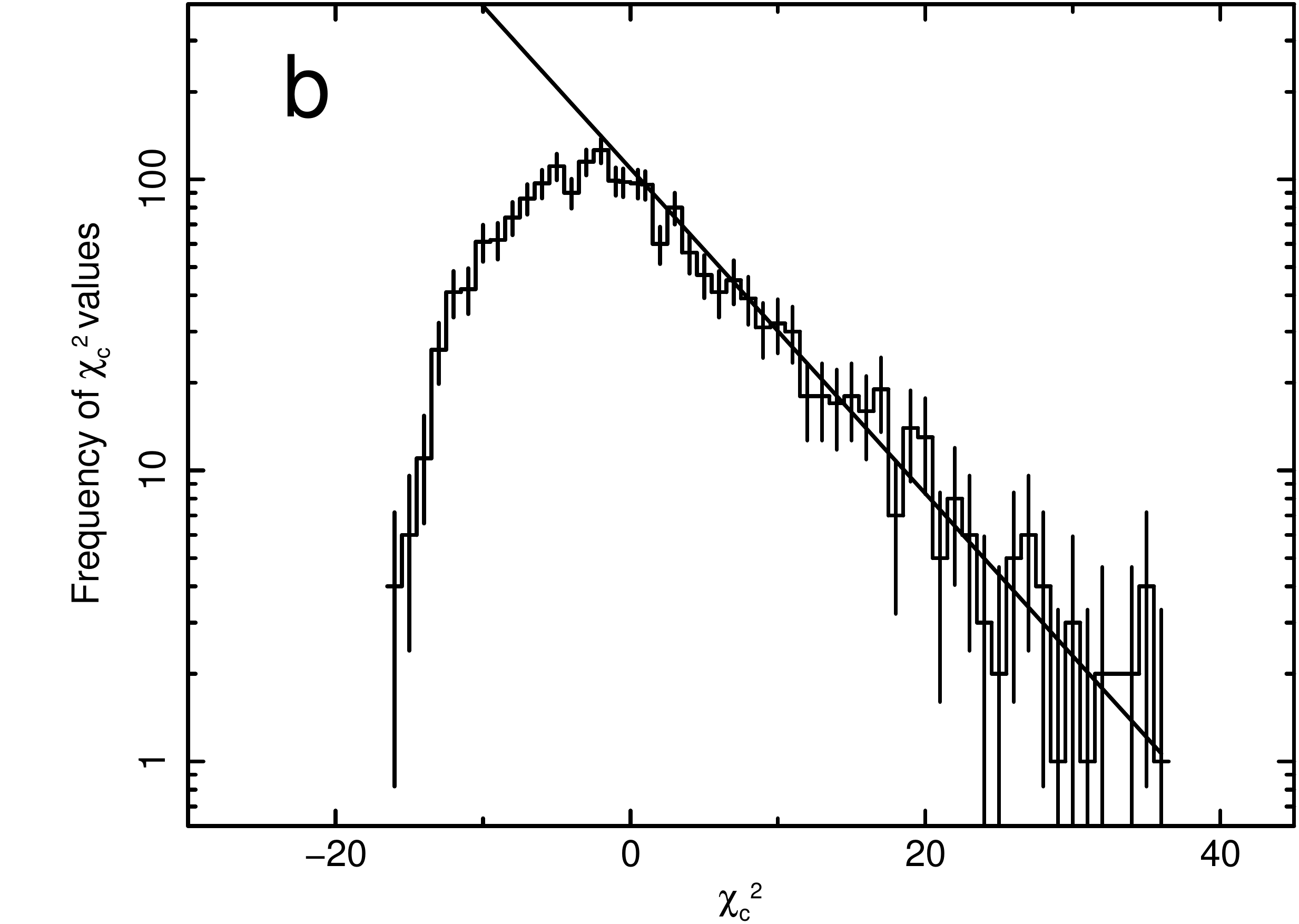}}
%\vspace{0.5truecm}
\centerline{\includegraphics[width=5.2cm,angle=270]{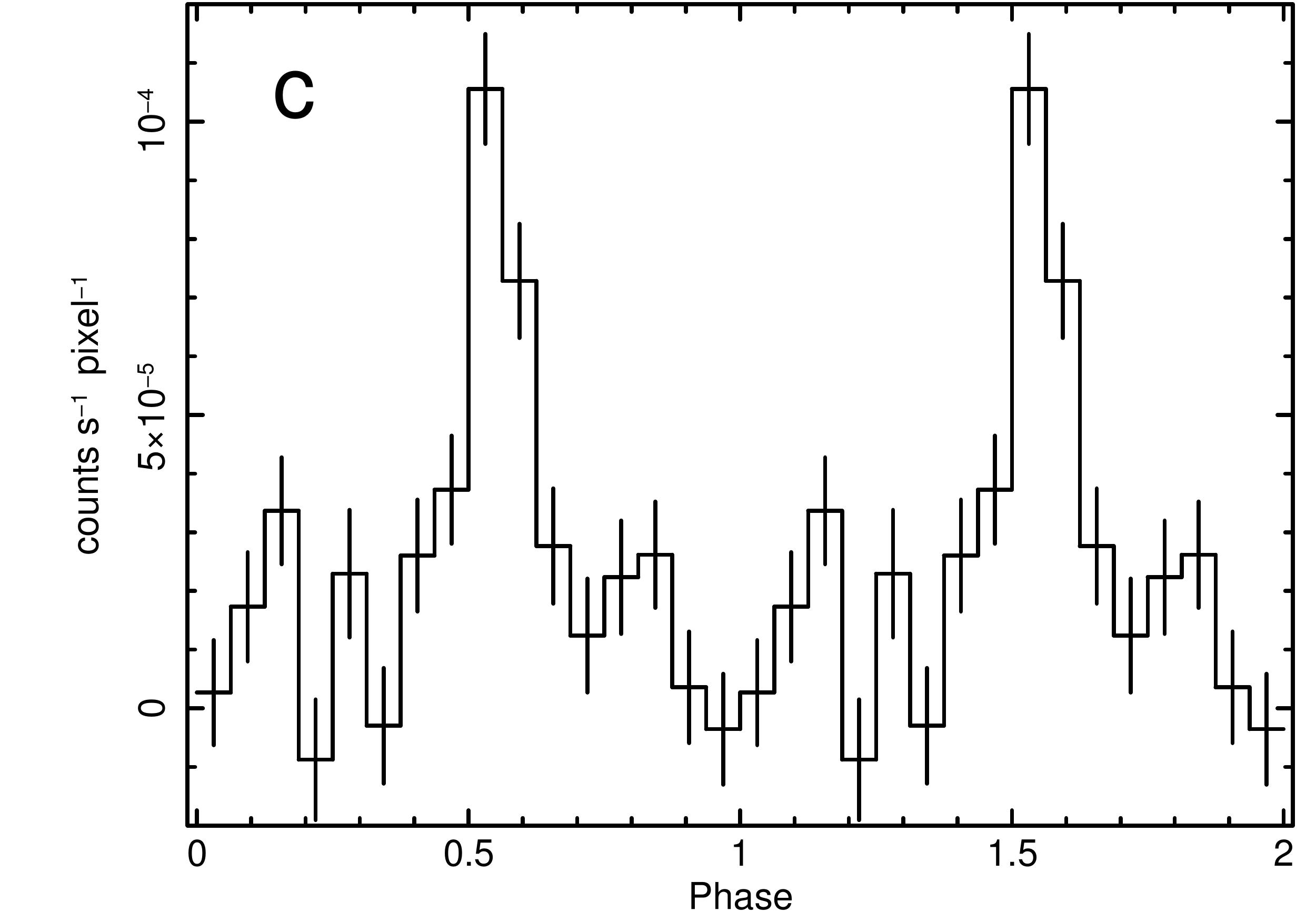}}
%\vspace{-2.5truecm}
\caption[]{{\bf  a}: Periodogram  of  \sw-BAT  (15--45\,keV) data  for
  J194221.  {\bf b}:  Distribution of the $z$ values  derived from the
  $\chi^2$ periodogram.   The positive tail beyond  $z$=10 is modelled
  with  an exponential  function.  {\bf  c}: Light  curve folded  at a
  period P$_0$=166.5\,days, with 16 phase bins. \label{period}}
\end{center}
\end{figure}

We performed  a timing  analysis of the  $Swift$/BAT data  searching for
long term periodic  modulations in the 15--45 keV  energy range during
the  enhanced activity  period from  2010.6 to  2014.5.  We  applied a
folding algorithm  to the barycentred  arrival times searching  in the
1--1000\,days time range with a  step of P$^{2}/(N \,\Delta T)$, where
P is the trial period, $N=16$ is  the number of profile phase bins and
$\Delta T$\,=\,124 Ms is the data time span.  The average rate in each
phase bin was  evaluated by weighting the rates by  the inverse square
of         the         corresponding        statistical         errors
\citep[see][]{cusumano10}.\\   Figure~\ref{period}   (a)   shows   the
periodogram with several features emerging.  The most prominent one is
at P$_0$\,=\,166.5\,$\pm$\,0.5 d  ($\chi^2$\,=\,172), where the period
and its error  $\rm \Delta P_0$ are  the centroid of the  peak and the
standard  deviation  obtained from  a  Gaussian  fit to  the  $\chi^2$
feature at P$_0$; other significant features at higher (lower) periods
corresponding to  multiples (sub-multiples) of P$_0$  are also clearly
detected in the periodogram.\\ The long-term variability of the source
causes the distribution of $\chi^2$ to deviate sharply both in average
and in fluctuation  amplitude from the behaviour expected  for a white
noise signal  dominated by  statistical variations.  As  a consequence
the $\chi^2$  statistics cannot be  applied, and for an  estimation of
the significance of  this feature, we applied  the following procedure
\citep[e.g.   see  also][]{dai11}: \\  i)  we  fitted the  periodogram
between 50  and 250  days (characterized by  a noise  level consistent
with the noise  level at P$_0$), excluding  opportune intervals around
the peaked features,  with a linear fit that describes  well the trend
of the $\chi^2$ values. The best fit function was then subtracted from
the $\chi^2$ to  obtain a new, flattened,  periodogram (hereafter, z);
the new $z$ value at $P_0$ is 148.2.\\ ii) We then built the histogram
of these  $z$ values, and  fitted the  positive tail of  the histogram
distribution  (beyond $z$\,=\,10)  with  an exponential  function, and  we
evaluated  the   integral,  $\Sigma$,   between  148.2   and  infinity
(normalized to the total area below the distribution).  The area below
the histogram  was evaluated summing  the contribution of  each single
bin from its  left boundary up to $z$\,=\,10 and  integrating the best-fit
exponential   function  beyond   $z$\,=\,10.    $\Sigma$  represents   the
probability of chance  occurrence of $z$ greater than  148.2.  \\ iii)
From the probability  of chance occurrence (2.2\,$\times$\,10$^{-9}$),
we evaluated the corresponding significance of the feature at P$_0$ in
units  of Gaussian  standard deviations  ($\sim$\,6.0\,$\sigma$).  The
light curve profile (Fig.~\ref{period}, panel  c) folded at P$_0$ with
T$_{\rm epoch}$ = 55828.91015625 MJD, is  characterized by a a peak of
intensity $\sim$\,4 times higher than the profile-averaged flux and by
a phase-span  of $\sim$\,1/8 of  the total period, corresponding  to a
duration of $\sim$\,20 days.

The three  $Swift$/XRT observations  fall at  phases 0.524,  0.543 and
0.553 and,  according to  the folded  profile, are  in phase  with the
emission peak.  We also searched  for the presence of shorter periodic
signals in the  $Swift$/XRT data, performing a timing  analysis on the
arrival  time of  the  source  events extracted  from  each  of the  3
$Swift$/XRT observations.  In order to avoid systematics caused by the
read-out time  in PC mode (characterized by a time resolution  bin of
$\delta$T$_{xrt}$\,=\,2.5073  s),   these  arrival  times  were  first
randomized  within $\delta$T$_{xrt}$.   Then  we  performed a  folding
search in  the range [$\delta$T$_{xrt}$  : 500] s, but  no significant
feature above the level of noise was detected.

%%%%%%%%%%%%%%%%%%%%%%%%%%%%%%%%%%%%%%%%%%%%%%%%%%%%%%%%%
\section{Spectral analysis} \label{spectral_analysis}
%%%%%%%%%%%%%%%%%%%%%%%%%%%%%%%%%%%%%%%%%%%%%%%%%%%%%%%%% 

We studied  the high-energy  BAT spectrum  collecting the  events only
after 2010.6 to  optimize the SNR, but keeping all  the available band
(15--150 keV energy  range) to better constrain the  cut-off energy of
the  spectrum.   We produced  a  phase-averaged  and a  phase-selected
spectrum in the  phase-interval 0.3--0.7 that corresponds  to the peak
of  the  folded  profile   of  Fig.~\ref{period}.  Both  spectra  were
accumulated  in  eight energy  channels  and  analysed using  the  BAT
redistribution   matrix    available   in   the $Swift$   calibration
database\footnote{\url{http://swift.gsfc.nasa.gov/docs/heasarc/caldb/swift/}}.
We fitted  together the two  spectra with  a simple power-law,  and we
obtained  a satisfactory  fit ($\chi^2/dof$\,=\,15/14)  with a  common
photon-index   ($\Gamma$\,=\,3.3\,$\pm$\,0.3)   and   free   to   vary
normalizations.  We  then adopted  a  bremsstrahlung  model having  an
exponential     high-energy    cut-off     and    we     obtained    a
$\chi^2/dof$\,=\,7/14 with a plasma temperature of 13$_{-3}^{+4}$ keV,
indicative of an  exponential decay of the spectrum  at high energies.
These results imply a factor $\sim$\,2 higher flux (20--100 keV energy
band)  at the  peak of  the folded  profile (1.7\,$\pm$\,0.3  $\times$
10$^{-11}$ erg cm$^{-2}$ s$^{-1}$)  with respect to the phase-averaged
flux  emission  (7.1\,$\pm$\,1.4  $\times$  10$^{-12}$  erg  cm$^{-2}$
s$^{-1}$), but undetected variability in the spectral shape.

We studied the  pointed $Swift$/XRT spectra in the  0.5--10 keV range,
while  we  used  the  0.3--0.7 phase-selected  $Swift$/BAT  spectrum  to
provide a  coverage of the hard  X-ray emission above 15  keV. The BAT
spectrum is averaged for many orbital  periods and its use in a common
fit with the  $Swift$/XRT spectra is helpful in giving  an estimate of
the broadband spectral shape  assuming negligible spectral variability
between the pointed and the long-term spectra.

\begin{figure}
\begin{center}
\centerline{\includegraphics[height=\columnwidth,angle=-90]{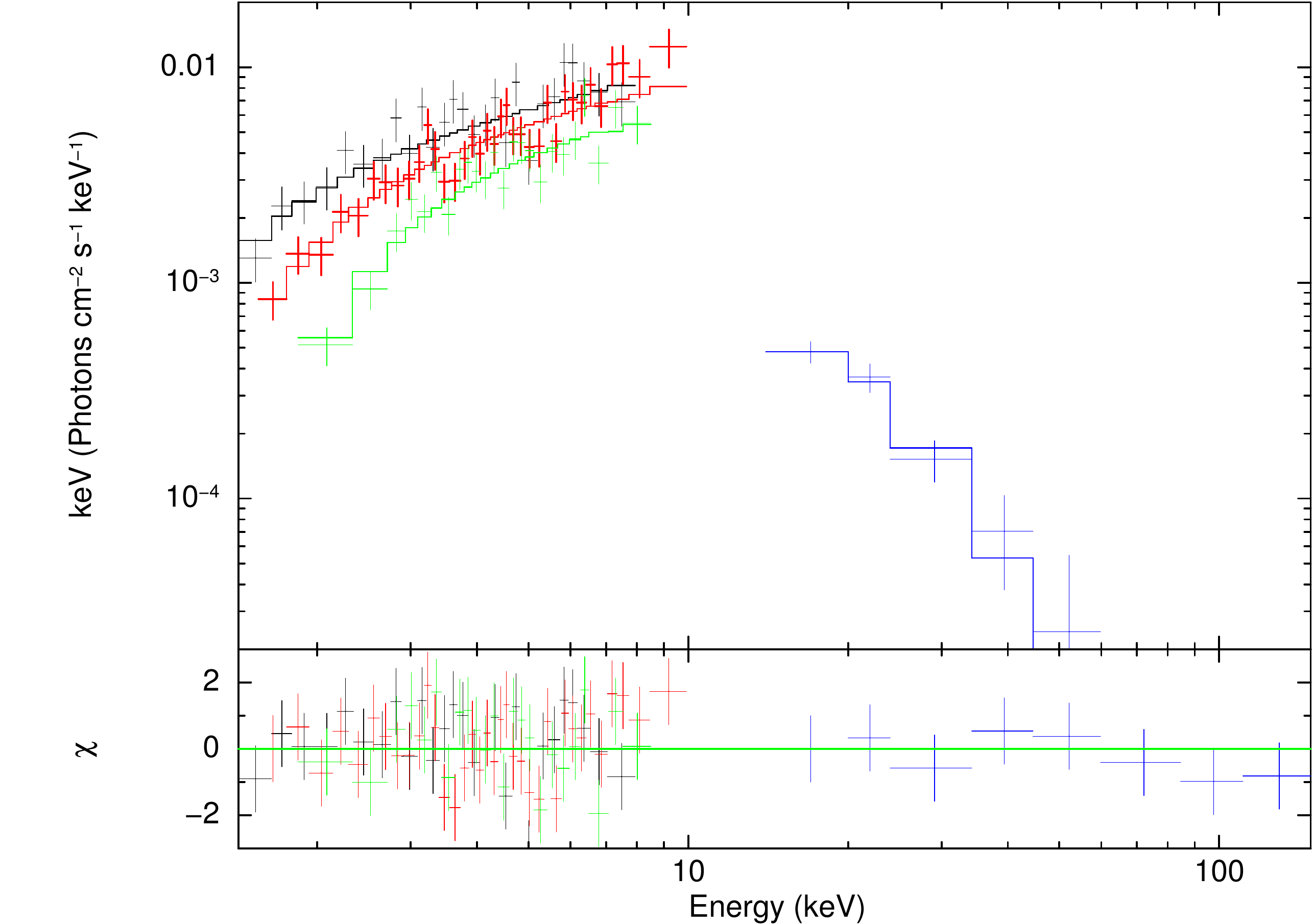}}
\caption{Data, best-fitting unfolded model (upper panel) and residuals
  in units  of $\sigma$ (lower  panel) for the combined  $Swift$/XRT and
  $Swift$/BAT  spectra, adopting  the model  of Table~\ref{fit}.  Black,
  red, green  and blue color for  the $Swift$/XRT Obs.ID 01,  02, 03 and
  $Swift$/BAT data, respectively.}
\label{spectrum}
\end{center}
\end{figure}

The  three  $Swift$/XRT  spectra showed  slightly  different  spectral
shapes  and  fluxes.   We  noted  that the  spectra  had  the  largest
discrepancy in the softest part of  the X-ray spectrum, and we took it
as  a   possible  indication  of   a  varying  local   column  density
($N_{\rm{H}}$).   Allowing for  different  values  of the  neutral
absorption  for  each  observation and  introducing  a  multiplicative
constant  to  account   for  the  different  fluxes,   we  obtained  a
satisfactory account of all the $Swift$/XRT spectra, consistent with a
hard power-law  of best-fitting  photon-index $\Gamma$\,=\,0.4.   When the
BAT spectrum  is added to the  fit, the simple power-law  model did no
longer provide  a good fit  to the  data (reduced $\chi^2$\,=\,3.0  for 88
dof), because  of the significant  steepening in the  $Swift$/BAT energy
range.   The  resulting  spectrum  was  much  better  fitted  (reduced
$\chi^2$\,=\,1.15,  for  87  dof)   with  a  cut-off  power-law  model
(\texttt{cutoffpl}  in  \textsc{XSPEC})  having  a  best-fitting  photon  index
$\Gamma$\,=\,-0.4\,$\pm$\,0.3      and      a      cut-off      energy      
$E_{\rm cut}=6.8_{-1.2}^{+1.6}~keV$.   Adopting a  model  with a  power-law
multiplied  by   a  high-energy  cut-off  with   an  e-folding  energy
($Exp[(E_{\rm cut}-E)/E_{\rm fold}])$, \texttt{highecut} model), we obtained a
reduced $\chi^2$  of 1.08 for 86  dof, that provides, however,  only a
marginal improvement (F-test chance improvement 4.4\%) with respect to
the     \texttt{cutoffpl}     model     (Fig.~\ref{spectrum}).      In
Table~\ref{fit},  we show  the  best-fitting  spectral parameters  and
error-bars  calculated  at  $\Delta \chi^2$\,=\,2.7  (90\%  confidence
level) for this final model.

\begin{table}
\begin{center}
\begin{tabular}{ll}
\hline
Parameter                              & Best-fitting value     \\ \hline \hline
\multicolumn{2}{c}{Common fitting values} \\ \hline
$\Gamma$                               & 0.4$\pm$0.2 \\
$E_{\rm cut}$ (keV)                    & 18$_{-10}^{+4}$ \\
$E_{\rm fold}$ (keV)                   & 7.3$_{-2.7}^{+3}$ \\ \hline
\multicolumn{2}{c}{Obs.ID 01} \\
\hline
N$_{\rm H}$ (10$^{22}$ cm$^{-2}$)  & 0.8$\pm$0.5 \\
$F_{0.5-10~\textrm{keV}}^{a}$  & 10$\pm$1.3  \\
\hline
\multicolumn{2}{c}{Obs.ID 02} \\
\hline
$N_{\rm H}$ (10$^{22}$ cm$^{-2}$)  & 1.7$\pm$0.6 \\
$F_{0.5-10~\textrm{keV}}^{a}$   & 9.0$\pm$0.6  \\
\hline
\multicolumn{2}{c}{Obs.ID 03} \\
\hline
$N_{\rm H}$ (10$^{22}$ cm$^{-2}$)  & 3.7$\pm$1 \\
$F_{0.5-10~\rm{keV}}^{a}$ & 6.6$\pm$0.6  \\
\hline
\multicolumn{2}{c}{BAT} \\
\hline
$C_{\rm BAT}$                          &  (8.6$_{-2.0}^{+20}$) $\times$ 10$^{-2}$\\
$F_{\rm 15-100~keV}^{a}$                   & 2.4$\pm$0.4\\
$\chi^2$ / dof                         & 93/86  \\ \hline
\multicolumn{2}{l}{$^a$ In units of 10$^{-11}$ erg cm$^{-2}$ s$^{-1}$}\\
\end{tabular}
\caption{Best-fitting spectral  parameters for the  combined $Swift$/XRT
  and Swif/BAT spectra adopting a  model of an absorbed power-law with
  a high-energy cut-off. We left the column density of the $Swift$/XRT
  data-sets free to vary, while the column density of the BAT spectrum
  is tied to the Obs.ID 01 value.}
\label{fit}
\end{center}
\end{table}

%%%%%%%%%%%%%%%%%%%%%%%%%%%%%%%%%%%%%%%%%%%%%%%%%%%%%%%%%%
\section{Discussion\label{discussion}}
%%%%%%%%%%%%%%%%%%%%%%%%%%%%%%%%%%%%%%%%%%%%%%%%%%%%%%%%% 

In this paper we presented the  first study on the timing and spectral
properties of the  X-ray source J1944221 exploiting  the data recorded
by  the  X-ray instruments  on  board  of  $Swift$.  The  source  showed
significant and  steady higher accretion  rates since 2011, and  it is
clearly  resolved  in hard  X-rays  up  to  $\sim$45 keV.  The  timing
analysis on the $Swift$/BAT survey revealed a periodic modulation with a
period of P$_0$\,=\,166.5$\pm$\,0.5 days, that we associate  to the orbital
period of  an X-ray  binary system.   The profile  of the  light curve
folded at  P$_0$ shows a  peak in emission  lasting $\sim$ 1/8  of the
period. The folded  profile is consistent with a  scenario of enhanced
accretion from  a compact object  close to the periastron  passage, as
typical for the class of  Be X-ray binary systems.  Our interpretation
agrees with \citet{masetti12atel}, that on the base of the presence of
strong  H-alpha  line  in  the  optical  spectrum,  first  proposed  a
high-mass  X-ray identification  for  the J194221.\\  We analysed  the
broad-band X-ray spectrum of J194221 using the XRT pointed observation
data in the soft X-ray band and the BAT survey data.  The spectrum can
be well modelled with a hard ($\Gamma$\,=\,0.4) absorbed power-law with an
energy  cut-off  $E_{\rm cut}$ at  $\sim$\,18  keV  and a  folding  energy
$E_{\rm fold}$ $\sim$\,7.3 keV.  We  found evidence of a  possible varying
local  neutral absorber  in the  $Swift$/XRT spectra.   Among the  three
examined observations,  the lowest column density  value is consistent
with the  Galactic value in  the direction of the  source \citep[$1.09
  \times  10^{22}$  cm$^{-2}$][]{dickey90},  while the  highest  value
indicate  moderate  local absorption.  The  hard  X-ray spectrum,  the
high-energy folding  and cut-off energies,  and the variations  in the
equivalent  hydrogen  column  observed  in  close  (within  few  days)
observations  at  the  putative  periastron passage  are  all  typical
spectral  characteristics of  the Be  X-ray binary  class and  further
support our identification of the source  as an active Be X-ray binary
\citep[see also for similar interpretations][]{laparola13, laparola14,
  cusumano13}.
\section*{Acknowledgements}
This research has made use of data and/or software provided by the High Energy Astrophysics Science Archive Research Center (HEASARC), which is a service of the Astrophysics Science Divi- sion at NASA/GSFC and the High Energy Astrophysics Division of the Smithsonian Astrophysical Observatory.\\
This work has been supported by ASI grant I/011/07/0.

\bibliographystyle{mn2e}
\bibliography{refs}
\label{lastpage}
\end{document}